\documentclass{moriond}
\usepackage{tikz}
\usepackage{calc}
\usetikzlibrary{shapes,arrows}
\usetikzlibrary{fit,calc}
\usetikzlibrary{positioning}

\newcommand\AliceApprovedLabel[4][1.0]{%
    \begin{scope}[x={($ (#2.south east) - (#2.south west) $ )},y={( $ (#2.north west) - (#2.south west)$ )}, shift={(#2.south west)}]
        \node[anchor=south west, inner sep={0.005 * #1 * \textwidth}] at (0, 0) {\fontfamily{pcr}\selectfont\resizebox{\textwidth*\real{#3}*\real{#1}}{!}{\textcolor[rgb]{0.698,0.698,0.698}{\textbf{#4}}}};
    \end{scope}
}
\newcommand\AlicePreliminary[3][1.0]{%
    \node[anchor=south west,inner sep=0] (image) at (0,0) {#2};
    \AliceApprovedLabel[#1]{image}{0.185}{#3}
}
\newcommand\AlicePreliminaryStandalone[3][1.0]{%
  \begin{tikzpicture}
    \AlicePreliminary[#1]{#2}{#3}
  \end{tikzpicture}
}

\usepackage[style=trad-unsrt,doi=false]{biblatex}
\DeclareFieldInputHandler{title}{}
\renewbibmacro{in:}{}
\AtBeginBibliography{\small}

\def\be{\begin{equation}}
\def\ee{\end{equation}}
\def\bea{\begin{eqnarray}}
\def\eea{\end{eqnarray}}

\usepackage{amsmath}

\newcommand{\pp}{\ensuremath{\text{p\kern-0.05em p}}}

\newcommand{\PbPb}{\ensuremath{\mbox{Pb--Pb}}}

\newcommand{\sqrtsNoNN}{\ensuremath{\sqrt{s}}}

\newcommand{\figRef}[1]{Fig.~\ref{#1}}

\newcommand{\GeVc}{\ensuremath{\text{GeV}\kern-0.05em/\kern-0.02em c}}

\newcommand{\pT}{\ensuremath{p_{\text{T}}}}

\newcommand{\pTJet}{\ensuremath{p_{\text{T,jet}}}}

\newcommand{\pTJetCh}{\ensuremath{p_{\text{T,jet}}^{\text{ch}}}}

\newcommand{\zg}{\ensuremath{z_{\text{g}}}}
\newcommand{\zcut}{\ensuremath{z_{\text{cut}}}}
\newcommand{\Rg}{\ensuremath{R_{\text{g}}}}
\newcommand{\nsd}{\ensuremath{n_{\text{SD}}}}
\newcommand{\kT}{\ensuremath{k_{\text{T}}}}

\newcommand{\deltaR}{\ensuremath{\Delta R}}

\newcommand{\dZero}{\ensuremath{\text{D}^{0}}}

\newcommand{\lnkT}{\ensuremath{\ln(\kT{})}}
\newcommand{\lnDeltaR}{\ensuremath{\ln(R/\deltaR{})}}

\newcommand{\Photo}{}

\begin{document}
\vspace*{4cm}
\title{QCD dynamics studied with jets in ALICE}

\author{Raymond Ehlers on behalf of the ALICE Collaboration}

\address{Physics Division, Oak Ridge National Laboratory,\\
Oak Ridge, TN 37831, USA}

\maketitle\abstracts{
Precise measurements and calculations of the internal structure of hadronic jets produced in high energy proton or lead collisions have become a prominent research area in recent years. Jet substructure provides information about quantum chromodynamics (QCD) and plays an important role in the study of the evolution of the quark-gluon plasma. The ALICE experiment is uniquely suited to provide insight into the smallest splitting angles due to high efficiency in the reconstruction of charged particles.
In this proceeding, we present an overview of recent ALICE results on jet substructure in pp collisions involving measurements of generalized angularities of groomed and inclusive jets, a new double-differential measurement of the Lund jet radiation plane for jets with a transverse momentum ($p_{\mathrm{T}}$) between 20 and 120 \GeVc{}, the first direct measurement of the dead-cone effect, and substructure measurements of heavy flavor tagged jets. These latest results provide new insights into the jet evolution by comparing to various theoretical predictions.}

\hypertarget{introduction}{%
\section{Introduction}\label{introduction}}

Collimated sprays of particles known as jets provide a variety of
opportunities to test QCD dynamics. As aggregate objects, the overall
jet structure and internal substructure are sensitive to the evolution
of the jet and its splittings. Such splittings contain a wealth of
information complementary to the overall jet properties, presenting the
opportunity to make stringent and detailed tests of QCD predictions.

In order to investigate and test QCD, ALICE \cite{Abelev:2014ffa} has
made a variety of such jet structure and substructure measurements.
ALICE is particularly well suited for these measurements due to the
precise charged-particle tracking in the central barrel, based around
the Inner Tracking System and Time Projection Chamber. This enables the
measurement of small splitting angles at high efficiency. The
electromagnetic calorimeter (EMCal) allows for a more complete
measurement of the jet energy, but with more limited acceptance.

Jets are reconstructed for a variety of jet resolution parameters,
\(R\), using the anti-\(\kT{}\) algorithm provided in FastJet
\cite{Cacciari:2011ma}. These jets were reconstructed in \(\pp{}\)
collisions at \(\sqrtsNoNN{}\) = 5.02 and 13 TeV, which were recorded in
2016, 2017, and 2018. Two classes of jets are discussed:
charged-particle jets, where only charged particles are used for jet
reconstruction, and full jets, where both charged particles and EMCal
cluster information are incorporated. Jets are required to be contained
fully within the acceptance of the central barrel or EMCal,
respectively.

\hypertarget{systematic-study-of-jets-splittings-with-the-lund-plane}{%
\section{Systematic study of jets splittings with the Lund
plane}\label{systematic-study-of-jets-splittings-with-the-lund-plane}}

In order to characterize the full set of jet splittings, their
properties can be recorded in the Lund plane \cite{Dreyer:2018nbf},
which describes the jet splitting phase space. After \(R=0.4\)
charged-particles jets are reconstructed, the jet constituents are
reclustered using the Cambridge/Aachen algorithm, producing an angular
ordered tree of splittings, with each node connected to subjets. This
clustering is unwound, recording \(\lnDeltaR{}\) (where
\(\deltaR{}=\sqrt{(\Delta \eta)^{2} + (\Delta \varphi)^{2}}\)) and
\(\lnkT{}=\ln{(p_{\text{T}}^{\text{sublead}} \sin{\deltaR{}})}\) for
each splitting, and then proceeding to the next one, iteratively
following the harder subjet. This set of splittings contribute to the
measurement of the primary Lund plane density

\vspace{-0.15cm}

\[ \rho(\deltaR{}, \kT{}) = \frac{1}{N_{\text{jets}}}\frac{\text{d}^{2}n}{\text{d}\ln{(R/\Delta R)}\text{d}\ln{\kT{}}}. \]

\vspace{-0.1cm}

For this measurement, Bayesian iterative unfolding was utilized to
correct for detector effects \cite{Adye:1349242}. Unfolding was
performed in three dimensions for the first time in ALICE, incorporating
\(\pTJetCh{}\), \(\lnDeltaR{}\), and \(\lnkT{}\)
\cite{ALICE-PUBLIC-2021-002}. The measurement is corrected for the
purity and efficiency of non-uniquely matched subjets, and subjet
mismatches are corrected in the unfolding. The number of jets
normalization is determined through a separate one-dimensional
unfolding.

\begin{figure}[t]
    \centering
    \includegraphics[width=0.65\textwidth]{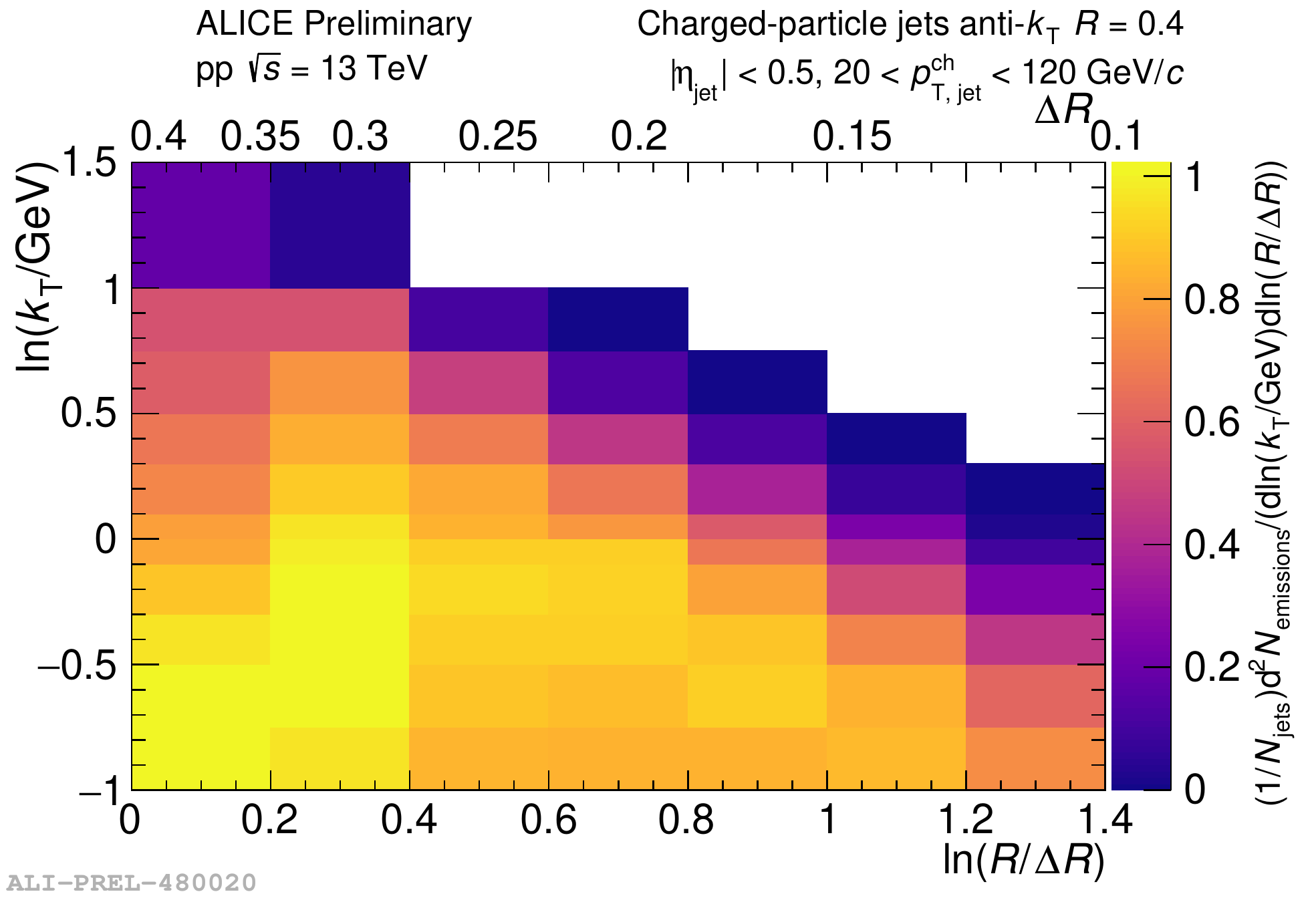}
    \caption{The fully corrected primary Lund plane density for $R = 0.4$ charged jets in $20 < \pTJetCh{} < 120$ \GeVc{}.}
    \label{fig:lundPlaneFullyCorrected}
\end{figure}

\begin{figure}[tb]
    \centering
    \includegraphics[width=0.4\textwidth]{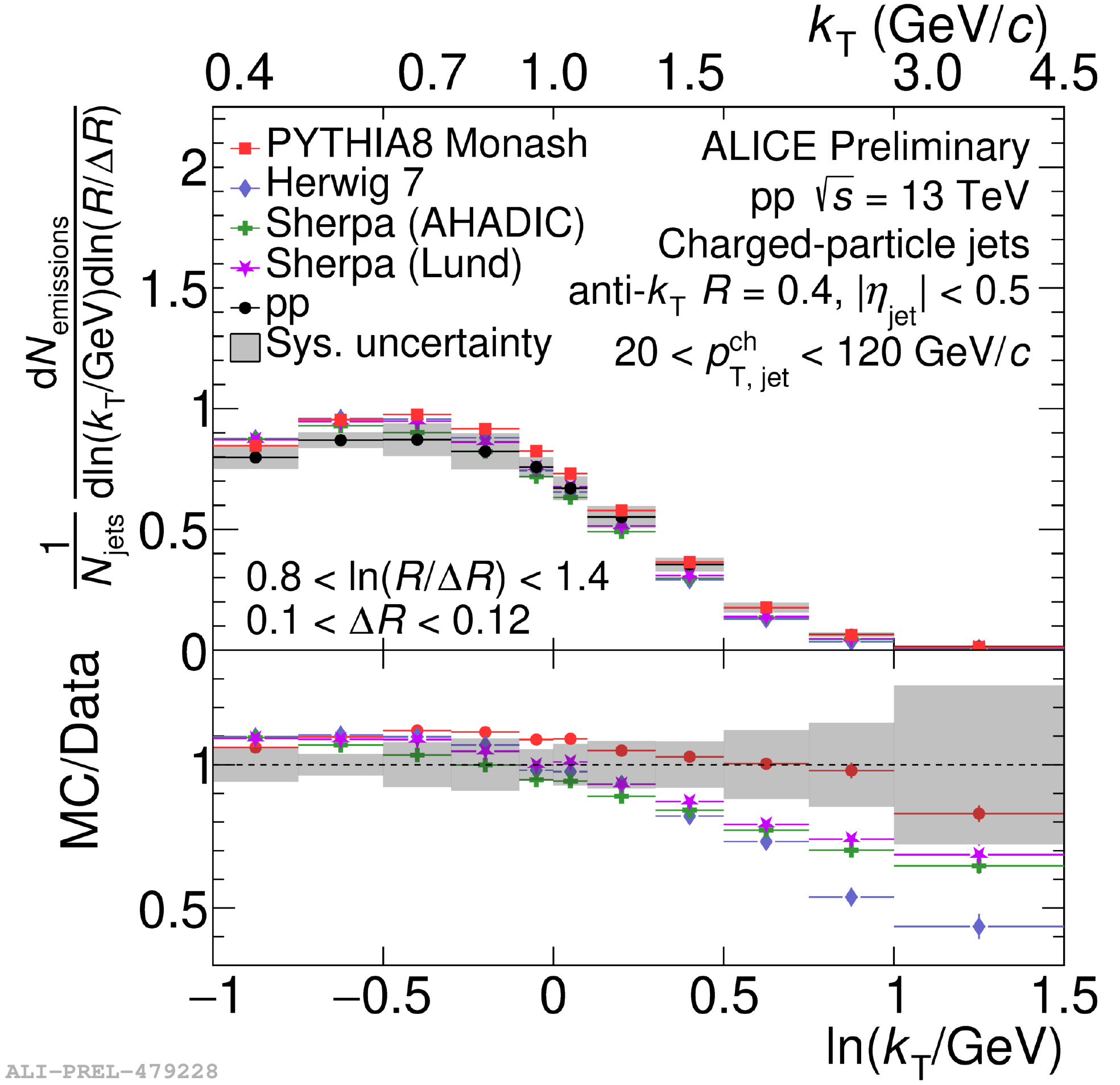}
    \includegraphics[width=0.4\textwidth]{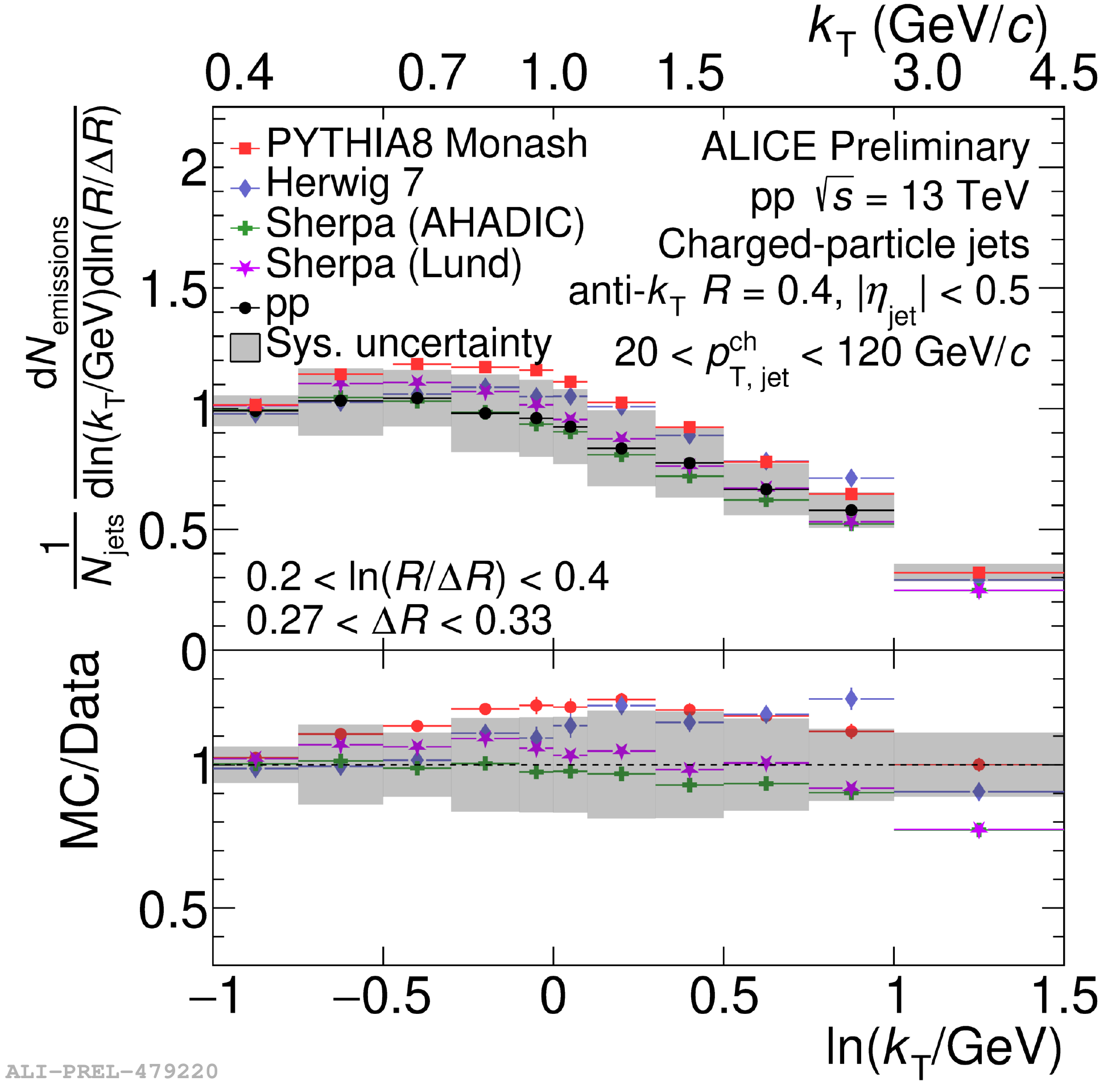}
    \caption{Selections of narrow (left) and wide (right) splittings from the primary Lund plane as a function of $\lnkT{}$ for $R = 0.4$ charged-particle jets.}
    \label{fig:lundPlaneNarrowVsWide}
\end{figure}

The fully corrected primary Lund plane density is shown in
\figRef{fig:lundPlaneFullyCorrected} for \(20 < \pTJetCh{} < 120\)
\GeVc{} measured in \(\pp{}\) collisions at 13 TeV. The most common
splittings are those at wide angles and low \(\kT{}\), steeply falling
with increasing \(\kT{}\). This measurement is at a substantially lower
\(\pTJet{}\) than the ATLAS measurement \cite{Aad:2020zcn}, providing
complementary information. In order to make more quantitative
statements, data are restricted into a variety of regions in phase space
and are compared to models, including PYTHIA8 Monash
\cite{Sjostrand:2007gs}, HERWIG 7 \cite{Bellm:2015jjp}, and SHERPA 2.28
\cite{Gleisberg:2008ta} with two different hadronization algorithms:
AHADIC and Lund. A selection of these comparisons are shown in
\figRef{fig:lundPlaneNarrowVsWide}, illustrating the different behaviors
for narrow vs wide splittings. Most of the models show some disagreement
with the data for narrow splittings at high \(\kT{}\), while the models
are able to describe the wider splittings within 10--20\%.

\hypertarget{observation-of-the-dead-cone}{%
\section{Observation of the Dead
Cone}\label{observation-of-the-dead-cone}}

Further tests of QCD are possible by focusing on the substructure of
charm quark tagged jets. Low energy splittings containing a \(\dZero{}\)
meson introduce mass effects from the charm quark, namely the
suppression of splittings at small angles due to the QCD dead cone
\cite{Dokshitzer:2001zm}.

\begin{figure}[tb]
    \centering
    \includegraphics[width=0.4\textwidth]{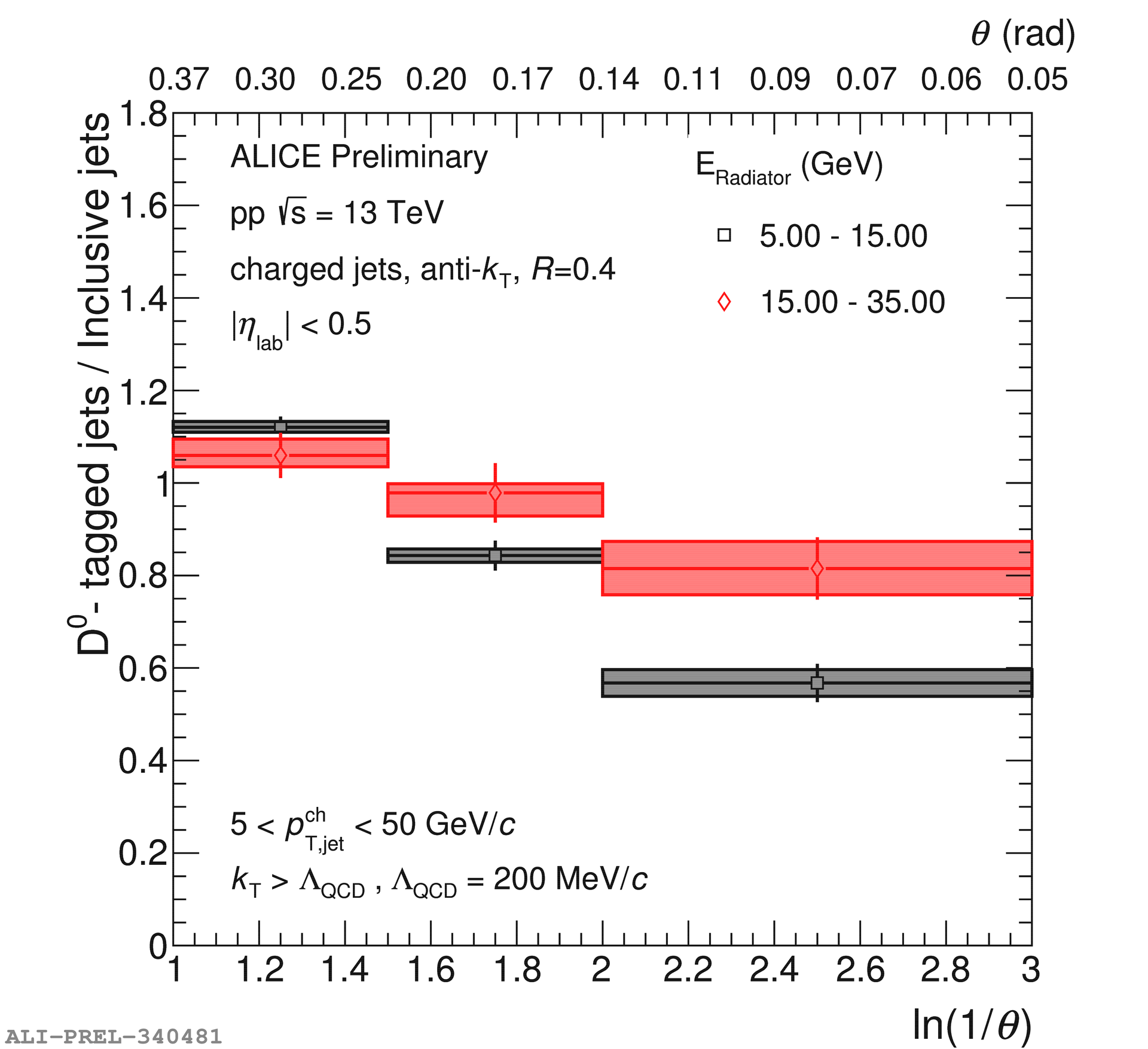}
    \includegraphics[width=0.4\textwidth]{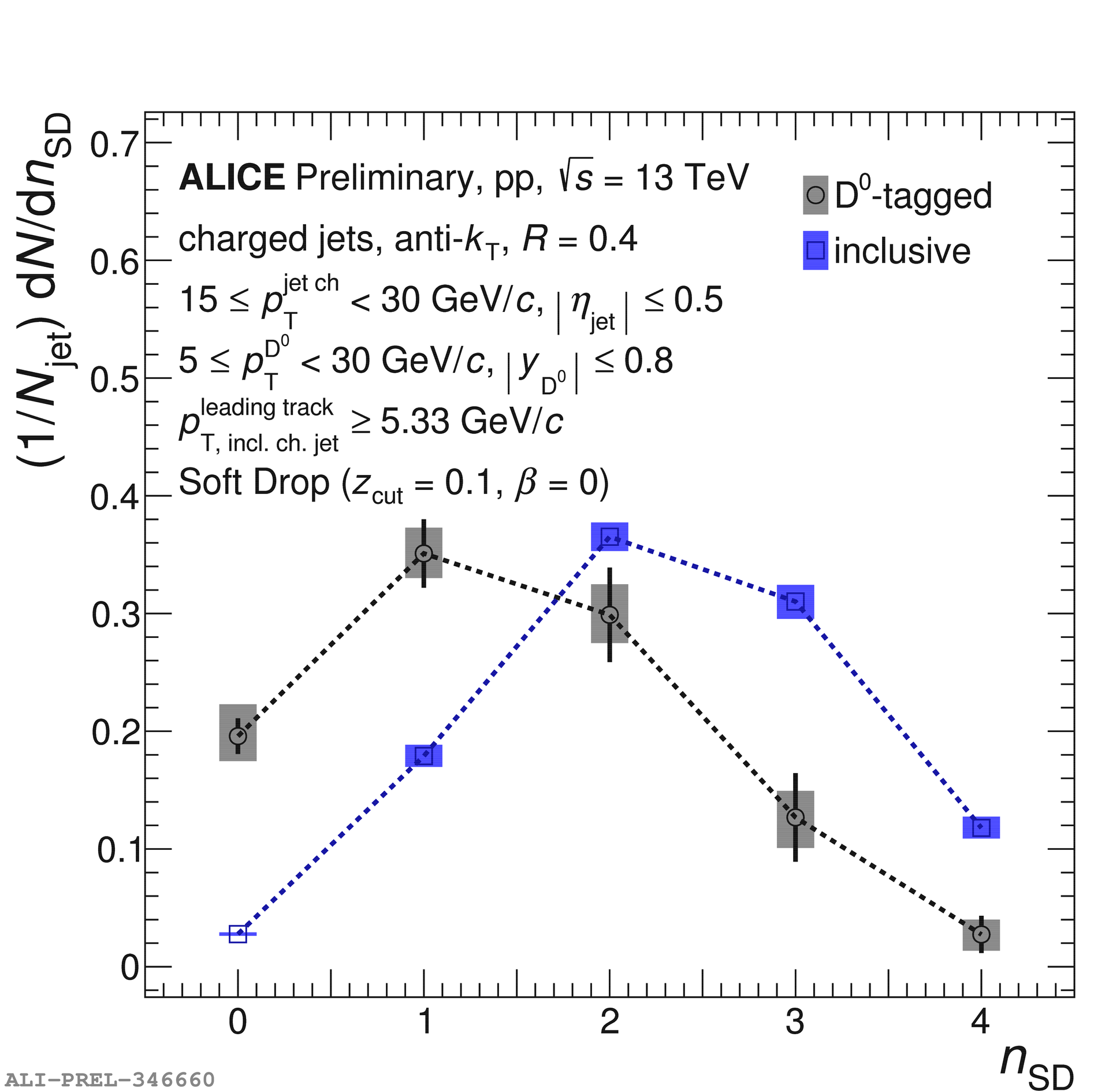}
    \caption{Angular separation of $\dZero{}$-tagged subjets compared to inclusive jets as a function of the incoming energy of the splitting (left) and
    the number of iterative splittings which pass the Soft Drop condition (right) $R = 0.4$ $\dZero{}$ tagged charged-particle jets in pp collisions at $\sqrtsNoNN{} = 13$ TeV. Both show modifications of low $\pT{}$ jets consistent with the dead cone effect.}
    \label{fig:deadCone}
\end{figure}

To access this effect, ALICE has measured the inverse of the splitting
angle \(1/\theta\) for all splittings in both \(\dZero{}\)-tagged and
inclusive charged-particles jets measured via the Lund plane in
\(\pp{}\) collisions at \(\sqrtsNoNN{} = 13\) TeV. In order to isolate
the impact of the dead cone, splittings are selected according to the
energy incoming to each splitting, \(E_{\text{radiator}}\). The ratio of
the rate of these splittings is shown on the left of
\figRef{fig:deadCone}. The class of lower \(E_{\text{radiator}}\)
splittings, which are more comparable to charm mass, are more suppressed
at small splitting angles compared to higher energies. This effect is
consistent with the QCD dead cone.

A similar effect can also be seen when selecting only a subset of
splittings, namely those which pass the Soft Drop
\cite{Larkoski:2014wba} condition, \(z_{\text{cut}} > 0.1\)
\cite{ALICE-PUBLIC-2020-002}. The number of splittings which pass this
condition, \(\nsd{}\), are compared for \(\dZero{}\)-tagged and
inclusive jets on the right of \figRef{fig:deadCone}. The reduced number
of splittings for \(\dZero{}\)-tagged jets is consistent with
expectations from color factors and the impact of the dead cone.

\hypertarget{selecting-specific-splitting-with-groomed-jet-substructure}{%
\section{Selecting specific splitting with groomed jet
substructure}\label{selecting-specific-splitting-with-groomed-jet-substructure}}

\begin{figure}[tb]
    \centering
    \includegraphics[width=0.3\textwidth]{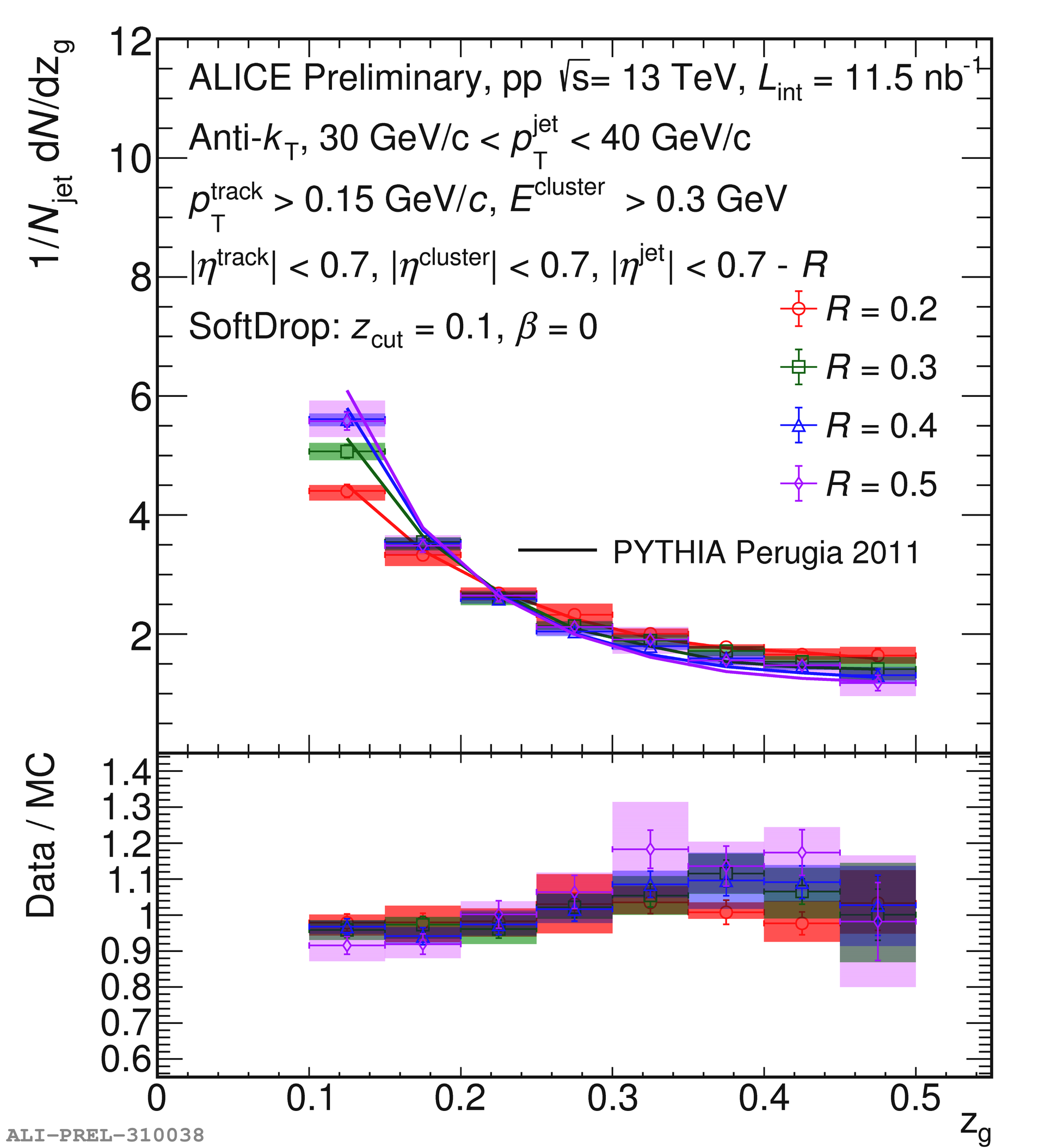}
    \AlicePreliminaryStandalone[0.425]{\includegraphics[width=0.425\textwidth]{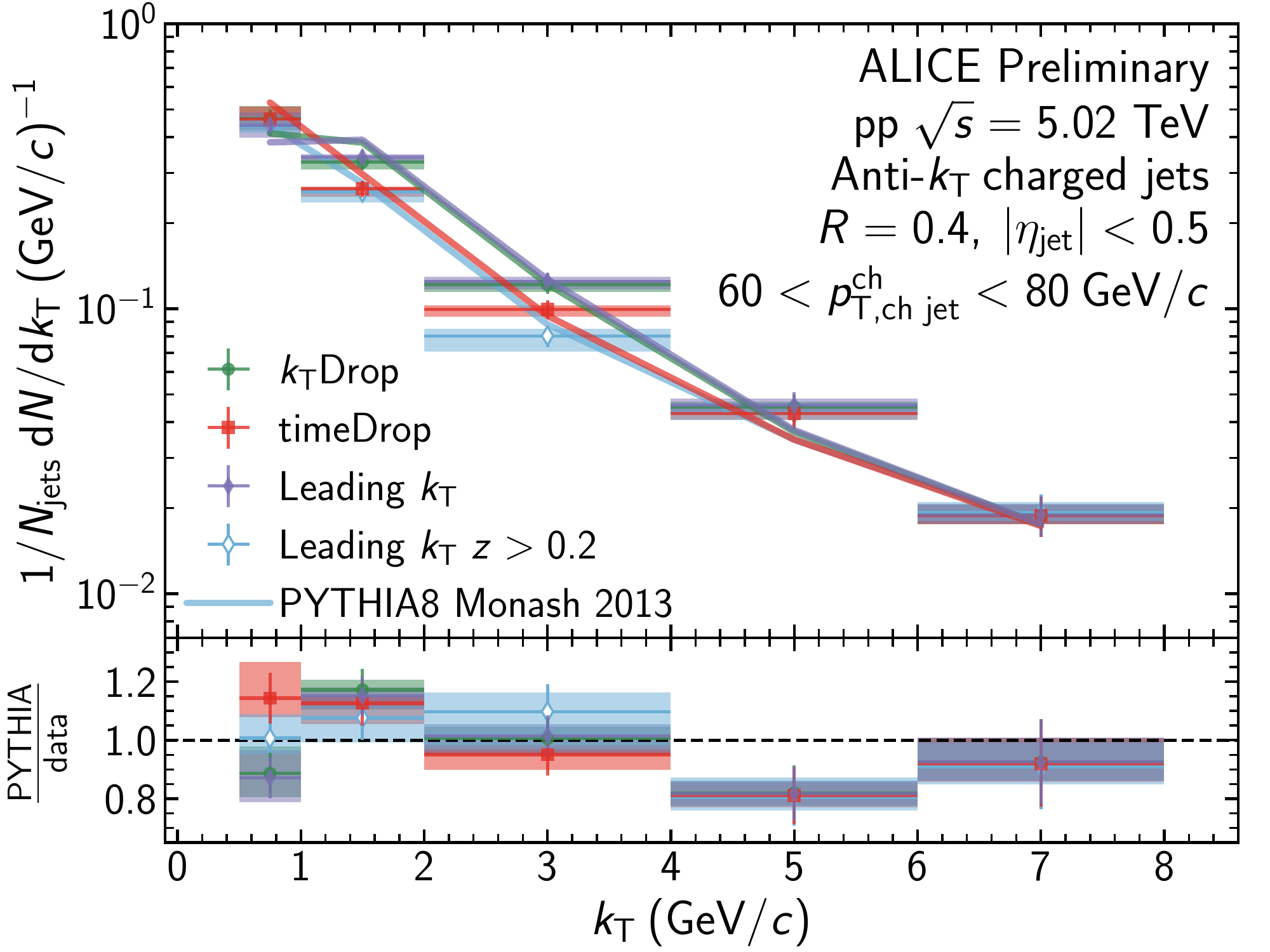}}{ALI-PREL-352234}
    \caption{Groomed shared momentum fraction via Soft Drop for $30 < \pTJet{} < 40$ \GeVc{} full jets in \pp{} collisions at $\sqrtsNoNN{}=13$ TeV (left), and groomed hardest $\kT{}$ splitting via a variety of grooming methods for $60 < \pTJetCh{} < 80$ \GeVc{} charged-particle jets in $\pp{}$ collisions at $\sqrtsNoNN{}=5.02$ TeV (right).}
    \label{fig:groomedSubstructure}
\end{figure}

In addition to \(\dZero{}\)-tagged jets, ALICE has performed detailed
studies of splittings in inclusive jets which are selected via a variety
of grooming methods. Systematic studies with Soft Drop
\cite{Larkoski:2014wba} using \(\zcut{} > 0.1\) have been performed for
\(R=\) 0.2-0.5 full jets measured in \(\pp{}\) collisions at
\(\sqrtsNoNN{} = 13\) TeV. A selection of these results at low
\(\pTJet{}\) are shown on the left of \figRef{fig:groomedSubstructure},
indicating that the shared momentum fraction is more symmetric for
smaller \(R\) jets. This \(R\) dependence disappears at high
\(\pTJet{}\) (\textgreater{} 160 \GeVc{}).

ALICE has also measured additional grooming methods, including the first
measurements using Dynamical Grooming \cite{Mehtar-Tani:2019rrk},
studying \(\Rg{}\), \(\zg{}\), and hardest \(\kT{}\) for \(R = 0.4\)
charged jets. The hardest \(\kT{}\) splittings are shown for a selection
of grooming methods on the right of \figRef{fig:groomedSubstructure},
illustrating the convergence of the selected splittings at high
\(\kT{}\) regardless of the method. Recent analytical calculations
\cite{Caucal:2021bae} show that the inclusion of non-perturbative
contributions is essential for describing the data.

\hypertarget{conclusion-and-outlook}{%
\section{Conclusion and outlook}\label{conclusion-and-outlook}}

ALICE has measured a wide variety of jet substructure measurements in
\(\pp{}\) collisions at \(\sqrtsNoNN{}\) = 5.02 and 13 TeV, including
the Lund plane for inclusive charged jets, \(\dZero{}\)-tagged jet
substructure, and groomed jet substructure for inclusive
charged-particle and full jets. Beyond these proceedings, relevant ALICE
measurements include full jet cross section ratios
\cite{Acharya:2019jyg} and systematic studies of charged-particle jet
angularities. Taken together, these observables provide substantial
opportunity to test QCD dynamics. Looking ahead, the Lund plane analysis
is a blueprint for future measurements, such as utilizing heavy flavor
tagged jets. These results will also serve as a baseline for comparison
with measurements in \(\PbPb{}\) collisions.

\vspace{-0.1cm}

\printbibliography

@article{Dokshitzer:2001zm,
	Author = {Dokshitzer, Yuri L. and Kharzeev, D. E.},
	Doi = {10.1016/S0370-2693(01)01130-3},
	Journal = {Phys. Lett. B},
	Pages = {199--206},
	Reportnumber = {LPT-ORSAY-01-58, BNL-NT-01-9},
	Title = {{Heavy quark colorimetry of QCD matter}},
	Volume = {519},
	Year = {2001}}

@article{ALICE-PUBLIC-2020-002,
	Author = {{{ALICE Collaboration}}},
	Collaboration = {ALICE Collaboration},
	Journal = {ALICE-PUBLIC-2020-002},
	Title = {{Groomed jet substructure measurements of charm jets tagged with $\rm{D}^{0}$ mesons in pp collisions at $\sqrt{s}$ = 13 TeV}},
	Url = {https://cds.cern.ch/record/2719005}}

@article{Bellm:2015jjp,
	Author = {Bellm, Johannes and others},
	Doi = {10.1140/epjc/s10052-016-4018-8},
	Journal = {Eur. Phys. J. C},
	Number = {4},
	Pages = {196},
	Reportnumber = {CERN-PH-TH-2015-289, MAN-HEP-2015-15, IFJPAN-IV-2015-13, KA-TP-18-2015, DCPT-15-142, MCNET-15-28, IPPP-15-71, HERWIG-2015-01},
	Title = {{Herwig 7.0/Herwig++ 3.0 release note}},
	Volume = {76},
	Year = {2016}}

@article{Gleisberg:2008ta,
	Author = {T. Gleisberg and others},
	Doi = {10.1088/1126-6708/2009/02/007},
	Journal = {JHEP},
	Pages = {007},
	Reportnumber = {FERMILAB-PUB-08-477-T, SLAC-PUB-13420, ZU-TH-17-08, DCPT-08-138, IPPP-08-69, EDINBURGH-2008-30, MCNET-08-14},
	Title = {{Event generation with SHERPA 1.1}},
	Volume = {02},
	Year = {2009}}

@article{Sjostrand:2007gs,
	Author = {Torbjorn Sjostrand  and others},
	Doi = {10.1016/j.cpc.2008.01.036},
	Journal = {Comput. Phys. Comm.},
	Pages = {852--867},
	Reportnumber = {CERN-LCGAPP-2007-04, LU-TP-07-28, FERMILAB-PUB-07-512-CD-T},
	Title = {{A Brief Introduction to PYTHIA 8.1}},
	Volume = {178},
	Year = {2008}}

@article{ALICE-PUBLIC-2021-002,
	Author = {{{ALICE Collaboration}}},
	Collaboration = {ALICE Collaboration},
	Journal = {ALICE-PUBLIC-2021-002},
	Title = {{Physics Preliminary Summary: Measurement of the primary Lund plane density in pp collisions at $\sqrt{s} = \rm{13}$ TeV with ALICE}},
	Url = {https://cds.cern.ch/record/2759456}}

@article{Dreyer:2018nbf,
	Author = {Dreyer, Fr\'ed\'eric A. and Salam, Gavin P. and Soyez, Gr\'egory},
	Doi = {10.1007/JHEP12(2018)064},
	Journal = {JHEP},
	Pages = {064},
	Reportnumber = {CERN-TH-2018-151},
	Title = {{The Lund Jet Plane}},
	Volume = {12},
	Year = {2018}}

@article{Acharya:2019jyg,
	Author = {{{ALICE Collaboration}}},
	Collaboration = {ALICE},
	Doi = {10.1103/PhysRevC.101.034911},
	Journal = {Phys. Rev. C},
	Number = {3},
	Pages = {034911},
	Reportnumber = {CERN-EP-2019-200},
	Title = {{Measurements of inclusive jet spectra in pp and central Pb-Pb collisions at $\sqrt{s_{\rm{NN}}}$ = 5.02 TeV}},
	Volume = {101},
	Year = {2020}}

@article{Caucal:2021bae,
	Archiveprefix = {arXiv},
	Author = {Caucal, Paul and Soto-Ontoso, Alba and Takacs, Adam},
	Eprint = {2103.06566},
	Primaryclass = {hep-ph},
	Title = {{Dynamical grooming meets LHC data}},
	Year = {2021}}

@article{Aad:2020zcn,
	Author = {{{ATLAS Collaboration}}},
	Collaboration = {ATLAS},
	Doi = {10.1103/PhysRevLett.124.222002},
	Journal = {Phys. Rev. Lett.},
	Number = {22},
	Pages = {222002},
	Reportnumber = {CERN-EP-2020-030},
	Title = {{Measurement of the Lund Jet Plane Using Charged Particles in 13 TeV Proton-Proton Collisions with the ATLAS Detector}},
	Volume = {124},
	Year = {2020}}

@article{Abelev:2014ffa,
	Author = {{{ALICE Collaboration}}},
	Collaboration = {ALICE},
	Doi = {10.1142/S0217751X14300440},
	Journal = {Int. J. Mod. Phys. A},
	Pages = {1430044},
	Reportnumber = {CERN-PH-EP-2014-031},
	Title = {{Performance of the ALICE Experiment at the CERN LHC}},
	Volume = {29},
	Year = {2014}}

@article{Adye:1349242,
	Author = {Adye, Tim},
	Journal = {{Unfolding algorithms and tests using RooUnfold}},
	Month = {May},
	Reportnumber = {arXiv:1105.1160},
	Title = {{Unfolding algorithms and tests using RooUnfold}},
	Url = {https://cds.cern.ch/record/1349242},
	Year = {2011}}

@article{Mehtar-Tani:2019rrk,
	Author = {Mehtar-Tani, Yacine and Soto-Ontoso, Alba and Tywoniuk, Konrad},
	Doi = {10.1103/PhysRevD.101.034004},
	Journal = {Phys. Rev. D},
	Number = {3},
	Pages = {034004},
	Title = {{Dynamical grooming of QCD jets}},
	Volume = {101},
	Year = {2020}}

@article{Larkoski:2014wba,
	Author = {Larkoski, Andrew J. and Marzani, Simone and Soyez, Gregory and Thaler, Jesse},
	Doi = {10.1007/JHEP05(2014)146},
	Journal = {JHEP},
	Pages = {146},
	Reportnumber = {MIT-CTP-4531, DCPT-14-24, IPPP-14-12},
	Title = {{Soft Drop}},
	Volume = {05},
	Year = {2014}}

@article{Cacciari:2011ma,
	Archiveprefix = {arXiv},
	Author = {Cacciari, Matteo and Salam, Gavin P. and Soyez, Gregory},
	Doi = {10.1140/epjc/s10052-012-1896-2},
	Journal = {Eur. Phys. J. C},
	Pages = {1896},
	Primaryclass = {hep-ph},
	Reportnumber = {CERN-PH-TH-2011-297},
	Title = {{FastJet User Manual}},
	Volume = {72},
	Year = {2012}}

\end{document}